# Symmetry energy of fragments produced in multifragmentation

D. Henzlova[1,a], A.S. Botvina[2], K.-H. Schmidt[1], V. Henzl[1,b], P. Napolitani[3], M.V. Ricciardi[1]

[1]*Gesellschaft für Schwerionenforschung, Planckstr. 1, 64291 Darmstadt, Germany*

[2]*Inst. for Nuclear Research, Russian Academy of Sciences, 117312 Moscow, Russia*

[3]*IPN Orsay, Université Paris-Sud 11, CNRS/IN2P3, 91406 Orsay cedex, France*

**Abstract**
Isospin properties of fragments measured in multifragmentation of $^{136}$Xe and $^{124}$Xe projectiles in mid-peripheral collisions with a lead target at 1 $A$ GeV were studied within the statistical approach describing the liquid-gas nuclear phase transition. By analyzing the isoscaling phenomenon and the mean $N$-over-$Z$ ratio of the fragments with $Z$ = 10-13 we have concluded that the symmetry energy of hot fragments produced in multifragment environment at subnuclear densities at high temperatures decreases in comparison with cold nuclei.

**Introduction**

In the recent years the interest in the isospin degree of freedom of the reaction products has considerably increased, motivated by the possibility to extract information on the symmetry energy of hot nuclei and nuclear matter during the liquid-gas phase transition. Such investigations are of particular interest for astrophysical applications, as previously discussed in [1,2,3]. It has been shown that the yield ratio of a given isotope produced in two reactions with different isospin asymmetries exhibits an exponential dependence on proton and neutron number, an observation known as isoscaling [4,5]. Isoscaling behavior has been identified in a variety of reaction mechanisms, including multifragmentation processes [6,7,8,9,10]. Based on the statistical interpretation of isoscaling, the coefficient of the symmetry-term in the nuclear mass can be extracted [8]. It can also be used for extraction of information concerning the density dependence of the symmetry energy [11,12].

**FRS experimental data**

In this work we present the analysis of fragments in the charge range $Z$ = 10-13, which may be associated with multifragmentation at high excitation energy, produced in the reactions of $^{136}$Xe+Pb and $^{124}$Xe+Pb at 1 $A$ GeV. We combine the isoscaling analysis with the experimentally determined <$N$>/$Z$ of the final residues to investigate the symmetry energy of hot primary fragments produced in the freeze-out volume at high excitation energy and subnuclear density. The residues produced in these reactions were identified with the use of the FRagment Separator (FRS) [13] at GSI, Darmstadt. The FRS is a high-resolution magnetic spectrometer with two magnetic stages separated by an intermediate dispersive image plane. All the residues passing through the FRS are identified in mass and nuclear charge by determining the mass-over-charge ratio $A/Z$ from the measured magnetic rigidity $B\rho$ and time-of-flight, and by deducing the atomic number $Z$ from the energy loss in an ionization chamber. The achieved resolving power in mass corresponds to $A/\Delta A \approx 400$ for all measured residues. The angular and momentum acceptance of the FRS correspond to 15 mrad and 3%, respectively. As a consequence, inclusive fragment yields in peripheral to mid-peripheral collisions are measured. Further details on the experimental approach may be found in [14].

---

[a] Present address: Los Alamos National Laboratory, Safeguards Science and Technology Group (N-1), Los Alamos, NM, 87545
[b] Present address: Massachusetts Institute of Technology, 77, Massachusetts Ave, Cambridge, MA, 02139

Fig.1 shows the mean neutron-to-proton ratio ($<N>/Z$) of the produced fragments for both projectiles. The final $<N>/Z$ above $Z \sim 45$ reveals a rather steep decrease with decreasing charge (increasing mass loss), corresponding to the dominating neutron evaporation from a rather low excited source. For lower charges (increasing excitation energy) the difference in the final $<N>/Z$ of the residues from the two xenon projectiles slightly decreases, but it survives over the whole charge range. This observation has been interpreted as indication for the break-up of a highly excited source [15]. Moreover, the residues with $Z < 15-20$ remain neutron rich in comparison with stable nuclei, which also suggests the presence of a nonevaporative process.

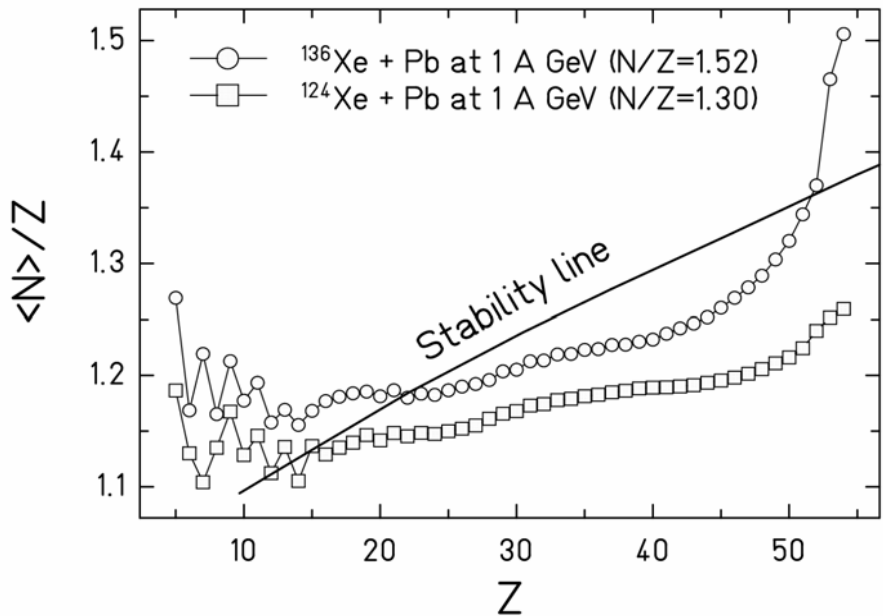

Fig. 1. Mean *N*-over-*Z* ratio of fragments produced in the reactions $^{136}$Xe+Pb and $^{124}$Xe + Pb in comparison with the valley of stability (solid line).

**Physical picture of the reaction**

We adopt the following physical picture of the fragment production: In peripheral nucleus-nucleus collisions after the fast dynamical stage, nuclei in a broad excitation-energy range are formed. It is the decay of these nuclei that produces the observed fragments. In the situation when an equilibrated source can be singled out in the reaction, statistical models have proven to be very successful. The most famous example of such a source is the compound nucleus introduced by Niels Bohr in 1936. The standard compound-nucleus approach is applied at low excitation energies when evaporation of nucleons and light charged particles, as well as fission are the dominating decay channels. However, this concept cannot be applied at high excitation energies, $E^* \geq 2-3$ MeV per nucleon, when the nucleus breaks up rapidly into many fragments. In the present work, we concentrate on the isotopic composition of smaller fragments, with $Z = 10-13$ which at these high excitation energies are predominantly produced in multifragmentation events characterized mainly by the emission of several fragments of similar size [16]. Such a situation is similar to conditions expected during the liquid-gas phase transition.

As a basis for our study we take the Statistical Multifragmentation Model (SMM), see review [17]. The model assumes statistical equilibrium at a low-density freeze-out stage, which may be reached at the time around 100 fm/c after the beginning of the reaction. These assumptions are consistent with model-independent analyses of experimental data [18,19,20,21] It considers all channels composed of nucleons and excited fragments taking into account the conservation of baryon number, electric charge and energy. Light nuclei with mass number $A \leq 4$ are treated as elementary particles with only translational degrees of freedom ("nuclear gas"). Nuclei with $A > 4$ are treated as heated liquid drops. In this way, one may study the nuclear liquid-gas coexistence in the freeze-out volume. The Coulomb interaction of all fragments is included using the Wigner-Seitz approximation. Different channels are generated by Monte Carlo sampling according to their statistical weights. An advantage of SMM is that it contains all break-up channels including the compound nucleus and one can study the competition between them. After the break-up, the Coulomb repulsion and the secondary deexcitation of primary hot fragments are taken into account.

In many experiments, the formation of an equilibrated source at high excitation energies was demonstrated (see e.g. [16,22,23,24,25,26]) and, as was shown, SMM is generally very successful in describing multifragmentation. The systematic study of such highly excited systems can bring important information about the parameters of the nuclear liquid-gas phase transition [15,27,28]. Since during this phase transition the uniform nuclear matter disintegrates into fragments of various masses, the properties of these fragments are of primary importance for probing the characteristics of the phase transition. We should note at this point that the question of the density and temperature dependence of the symmetry energy is often addressed in the case of a uniformly nuclear matter expanded without clusterization, see, e.g., ref.[29]. In the SMM we have another physical situation: The primary fragments are located in the freeze-out volume in the close vicinity of other nuclear species, that can be considered as a piece of nuclear matter at subnuclear density with clusters. In this respect, we analyze the matter after clusterization, and properties of nuclear clusters (fragments) inside this matter. Because of interaction with environment their properties should be quite different as compared to isolated cold fragments.

It is important for our present goal that the symmetry energy of hot fragments with mass number $A$ and nuclear charge $Z$ is parameterized in SMM as $E_{AZ}^{sym} = \gamma(A - 2Z)^2 / A$, where $\gamma$ is the symmetry-energy parameter. In order to fit the binding energies of isolated cold nuclei in the ground state, $\gamma = 25$ MeV should be considered. In mass formulas of cold nuclei this parametrization is associated with the bulk term of the symmetry energy. In order to improve the description of experimental masses one can introduce the surface term of the symmetry energy [30,31]. However, this kind of parameterization remains disputable even for cold isolated nuclei, since some modern formulas can provide even better description of nuclear masses, retaining only the bulk isospin term but with a special treatment of the shell effects [32]. It is well known that the shells are washed out with excitation energy [33], which is typical for multifragmentation. We believe also that the isospin properties of hot primary fragments may be different from the fragments in ground states because of the proximity to other nuclei. For this reason, their isospin properties may be different from fragments in ground states. There are theoretical calculations of multifragmentation reactions indicating that the bulk symmetry term alone is sufficient to describe isotope distribution of hot fragments [34]. Also there is the analysis of ALADIN experimental data obtained in fragmentation of relativistic spectators (similar to our reactions) which gives evidences that the surface symmetry term may disappear as the system enters multifragmentation regime [35]. One may attribute this effect to the remaining interaction between fragments inside the freeze-out volume. In view of a great uncertainty in understanding of this many-body process, we believe that the best strategy for investigating "in-medium" properties of hot nuclear fragments is to consider a minimum number of parameters which have a clear physical meaning, as the standard SMM parametrization used for the symmetry energy. Previously, the effects of modifications of volume and surface energy terms of primary fragments were investigated within SMM [35,50]. As was concluded that they do not affect the isospin properties of the fragments.

**Fragment isoscaling analysis**

For the purpose of investigating the symmetry-energy coefficient, we use the isoscaling approach, which is directly connected with the isotopic distribution of fragments. The isoscaling is more convenient for extracting the isospin properties averaged over all fragments than the original isotope distributions, since it allows for eliminating fluctuations of yields caused by structure effects [36]. The isoscaling concerns the production ratios $R_{21}$ for fragments with given neutron number $N$ and proton number $Z$ in reactions with different isospin asymmetry. It is constituted by their exponential dependence on $N$ and $Z$ according to:

$$R_{21}(N,Z) = Y_2(N,Z)/Y_1(N,Z) = C \cdot \exp(\alpha N + \beta Z) \qquad (1)$$

with three parameters, $C$, $\alpha$ and $\beta$. Here, $Y_2$ and $Y_1$ denote the yields from the more neutron-rich ($^{136}$Xe) and the more neutron-deficient ($^{124}$Xe) system, respectively.

Fig. 2 (upper panel) shows the isoscaling exhibited by fragments with $Z$=5-51 measured in the reactions of $^{124}$Xe and $^{136}$Xe+Pb. In the lower panel, the corresponding isoscaling coefficients $\alpha$, determined from best fit to the experimental data, are shown. One can see that for $Z \geq 40$ $\alpha$ decreases strongly with decreasing charge, which is consistent with the production of large fragments by an evaporation process at small excitation energy. For lower $Z$ the decrease is smoother with much lower values of $\alpha$. Fragments with $Z$ = 10-13 lead to $\alpha \approx 0.36$.

It was shown in [8] that the parameter $\alpha$ can be deduced within the statistical approach as:

$$\alpha \approx 4\frac{\gamma}{T}\Delta\left(\frac{Z^2}{A^2}\right) = 4\frac{\gamma}{T}\left(\frac{Z_1^2}{A_1^2} - \frac{Z_2^2}{A_2^2}\right) \qquad (2)$$

where $Z_1$, $A_1$ and $Z_2$, $A_2$ are the nuclear-charge and mass numbers of the two multifragmenting systems. The parameter $\alpha$ depends essentially only on the coefficient $\gamma$ of the symmetry term, on the temperature and on the isotopic compositions of the sources. With the use of this relation, the symmetry term may be extracted from the isoscaling parameter, provided the temperature and the isotopic composition of the sources are known.

Relations similar to equation (2) were also discussed in [7] within the EES model and for nuclear clusters produced in the AMD model [37] assuming that they reach the thermodynamical equilibrium after 300 fm/c. However, the physical meaning of this formula is different in those cases. In the EES relation (2) is deduced for the uniform nuclear matter of an expanding source, which evaporates cold fragments. This model can not be used for the interpretation of our data since it can not describe fragments with $Z > 9$. In the AMD model formula (2) was obtained for hot fragments separated, however, at later times, around 300 fm/c. This approach is more similar to ours, since it also addresses the properties of hot thermalized fragments. However, AMD fragments should be cooler than our fragments in the freeze-out volume, and equation (2) should have a different $\Delta(Z^2/A^2)$ factor as a consequence of the evolution of the fragments during their propagation from the freeze-out time of the order of 100 fm/c to 300 fm/c. We believe that our approach, which considers the statistical equilibrium in the freeze-out, has an important advantage. It allows for direct connection of our fragments with fragments produced at similar subnuclear densities in astrophysical conditions, for example, during collapse and explosion of massive stars, where the nuclear equilibrium is established [38].

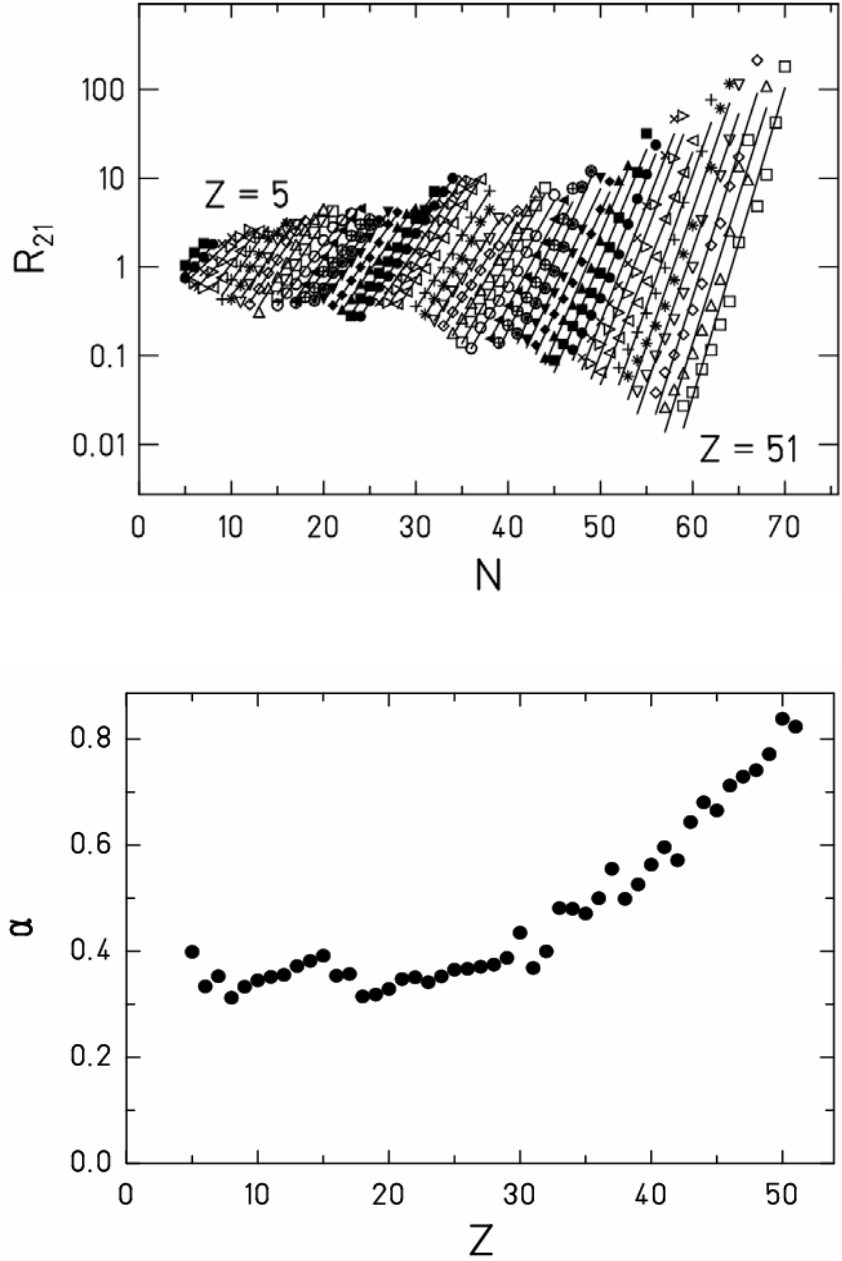

Fig. 2. The isoscaling (see eq. (1)) exhibited by fragments with $Z$=5-51 (upper panel), and the corresponding exponent $\alpha$ extracted from the experimental data (lower panel).

We compare our experimental value of $\alpha$ with results obtained in other experiments for the sources with similar isotopic content: In ref. [39], for the same $\Delta(Z^2/A^2)$ factor, the $\alpha$ determined for light fragments with $Z = 3$-7 decreases with increasing excitation energy from 0.38 to 0.24. In order to compare the data obtained for other sources we introduce the reduced quantity scaled by the isotopic composition of the investigated systems: $\alpha_{red} = \alpha/[4\cdot\Delta(Z^2/A^2)]$. According to equation (2) $\alpha_{red} = \gamma/T$. In our case $\alpha_{red} = 2.81$, in the case of [39] the corresponding $\alpha_{red}$ decreases from 2.97 to 1.88. For the experiment of ref. [7] we obtain $\alpha_{red} = 2.43$ for $Z$=1-8, and in the experiment of ref. [10] the value of $\alpha_{red}$ decreases from 4.19 to 1.69 with increasing excitation energy for $Z = 1$-5. The difference between these results may be partly explained by a slight variation of the temperatures reached in the reactions in case of production of fragments with different sizes. Another physical reason may be the variation of the symmetry coefficient $\gamma$ discussed in this paper.

To determine the temperature, we use the isospin-thermometer method, established in [15], where the two independent analyses with the SMM[17] and ABRABLA[40] codes were used to extract the temperature from the experimental data. We estimate the temperature for the case of the data in the present work, within the uncertainty of the method, as $T \approx 5$ MeV. This value is supported by other experimental investigations of the excited source of similar size as in the case of our experiments [41,42,43], as well as by the SMM code predictions of the temperature of the source producing the fragments in the charge range considered in our work [17,44]. According to these analyses, the temperature of the multifragmenting source ranges between $T \sim 4$-$6$ MeV. During dynamical stage of relativistic heavy-ion collisions fast protons and neutrons are knocked out approximately with equal probability, since the isospin-dependent effects are relatively small. As studied in [10] using INC and RBUU calculations, in the relativistic energy regime the factor $\Delta(Z^2/A^2) = (Z_1/A_1)^2 - (Z_2/A_2)^2$ may be approximated within few percent by the difference of the $Z/A$ of the initial nuclei. By applying the above formula to the experimentally determined $\alpha$ and considering the uncertainty in the temperature and $\Delta(Z^2/A^2)$ determination, we obtain an 'apparent' $\gamma_{ap} \approx 14\pm3$ MeV, that is essentially lower than the value of $\gamma = 25$ MeV. This result is in agreement with the analysis of the experimental isoscaling performed in [10], where, however slightly higher values of temperature and lower value of $\alpha$ (for the highest centrality bin) were reported than in the present work. It should be emphasized that lighter isotopes ($Z = 1$-$5$) are used to determine $\alpha$ in [10]. Isotopes investigated in our analysis are larger ($Z = 10$-$13$) and, therefore, they may be produced at lower temperatures. We should mention, however, that the above analysis is performed under assumption that the effect of secondary de-excitation processes on the $\alpha$ parameter is minimal [7,8].

**Influence of secondary de-excitation**

In the statistical approach, the formula (2) was obtained for the freeze-out conditions. In order to establish the connection between $\gamma$ of hot fragments and $\gamma_{ap}$ obtained for the observed cold fragments we should take into account the process of secondary deexcitation. Presently, there are various secondary-deexcitation codes combined with multifragmentation (e.g. [45,46]), which describe essentially a sequential evaporation for fragments with $Z \geq 10$. However, all these codes use standard mass formulas fitting masses of cold isolated nuclei. If hot fragments in the freeze-out configuration have smaller $\gamma$, their masses at the beginning of the secondary deexcitation will be different, and this effect should be taken into account in the evaporation process. Unfortunately, the magnitude of this effect has an uncertainty, since we do not know exactly at what stage of the secondary deexcitation the nuclei restore the experimental masses with the standard symmetry energy. It may happen at later times during the Coulomb propagation of the fragments from the freeze-out volume, when the fragments cool down sufficiently. However, it is well known that it should take place for isolated nuclei at $T \leq 1$ MeV [33].

We believe that we can estimate the effect of the symmetry-energy evolution during the sequential evaporation by the following phenomenological prescription (see also ref. [48]: For a given nucleus $(A,Z)$ evaporating the lightest particles (n,p,d,t,$^3$He,$\alpha$), we take liquid-drop masses adopted in the SMM for hot fragments, i.e. $m_{A,Z} = m_{ld}(\gamma)$, if the internal excitation energy of this nucleus is large enough $\xi = \beta \cdot E^*/A > 1$, [ $\beta = 1$ MeV$^{-1}$]. At lower excitation energies ($\xi \leq 1$) we adopt a smooth transition to standard experimental masses with shell effects ($m_{exp}$) with the following linear dependence:

$$m_{A,Z} = m_{ld}(\gamma)\cdot \xi + m_{exp} \cdot (1-\xi) \qquad (3)$$

The excitation energy is always determined from the energy balance taking into account the mass $m_{A,Z}$ at the given excitation. This mass correction was included in a new evaporation code developed on the basis of the old model [45], taking into account the energy, momentum, mass-number and charge conservations. We have checked that the new evaporation model with $\gamma = 25$ MeV leads to results very close to the standard evaporation [45] concerning the mean $N/Z$ ratio of cold fragments and their charge yield.

To estimate the influence of the secondary deexcitation we performed Markov-chain SMM calculations [47] for $^{136}$Xe and $^{124}$Xe projectile sources at excitation energies of $E^* =$ 4-6 $A$ MeV, which are expected in multifragmentation, where the fragments in the nuclear-charge range investigated in this work are predominantly produced. The density of the freeze-out was taken as $\rho = 0.3\,\rho_0$ (where $\rho_0 = 0.15$ fm$^3$ is the normal nuclear density). This version was adopted to take into account all finite-size effects. In order to check source-size effects on the results we performed similar calculations for sources with 60% of the projectile masses and with the same $Z/A$ ratios. We found that the isospin characteristics under study do not change significantly. As SMM input we varied $\gamma$ of the hot fragments in the range from 4 to 25 MeV, to see how the symmetry energy changes the isospin properties of the produced fragments. As was discussed in [48] the value of $\gamma$ only slightly influences the mean charge distributions of hot fragments, however, it considerably influences the isotopic distributions.

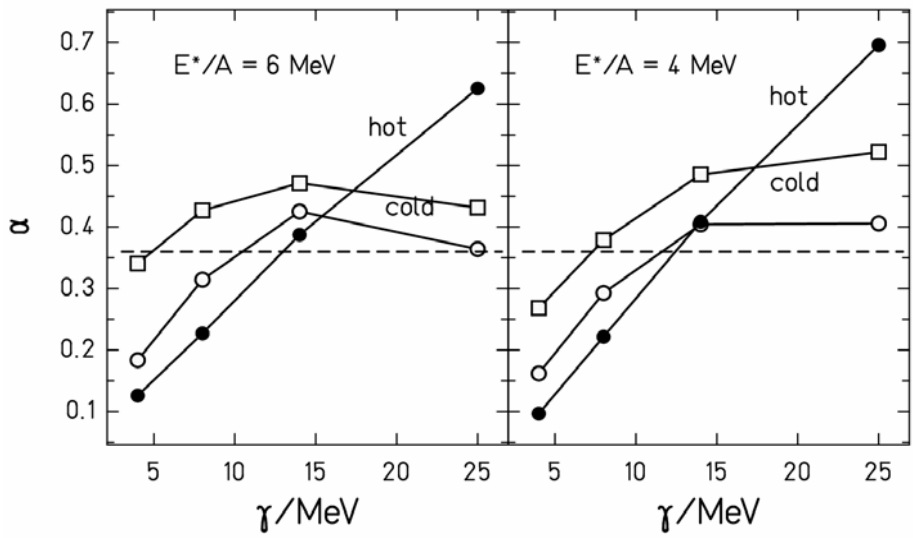

Fig. 3. The isoscaling parameter $\alpha$ versus the symmetry-energy coefficients $\gamma$ obtained from fragments with $Z$ = 10-13 in SMM calculations for $^{136}$Xe and $^{124}$Xe sources, and with excitation energies 6 $A$ MeV (left panel) and 4 $A$ MeV (right panel). Solid symbols represent primary hot fragments, empty symbols show final cold fragments; the new evaporation model (open circles), the old evaporation model (open squares). The dashed line shows $\alpha$ extracted from the experimental data.

Fig. 3 shows the calculated $\gamma$ dependence of the isoscaling parameter $\alpha$ extracted from the hot primary fragments in the freeze-out volume, and for the cold fragments after the secondary deexcitation. The $\alpha$-parameter for hot fragments exhibits a linear dependence on $\gamma$ expected also in the grand-canonical approach, see Eq. (2). The secondary evaporation pushes the isotopes towards the valley of stability, however, the final distributions still depend on the initial distributions of the hot fragments. One can see that the results also change depending on whether the symmetry energy evolves during the evaporation or not. Assuming $\gamma = 25$ MeV for hot fragments, the evaporation causes a slight broadening of the isotopic distributions with respect to the initial ones, and the resulting $\alpha$ is lowered. For smaller values of $\gamma$, however, the dominant effect is caused by the decay of the wings of the wider initial distribution, as a result the cold distributions are narrower and the corresponding $\alpha$ is larger. We demonstrate the two kinds of evaporation calculations, one is performed with the standard code [45], and the other one is based on the above described version taking into account the symmetry energy (mass) evolution during evaporation. The new model predicts final values of $\alpha$ much closer to the initial ones at smaller $\gamma$. The difference between the two versions of the calculation serves as a qualitative measure of the uncertainty expected in the secondary deexcitation. We observe that the experimentally determined value of α can be reproduced only by lower values of the γ coefficient, e.g. γ ≈ 5-8 MeV and 11-12 MeV with the old and new evaporation calculation, respectively.

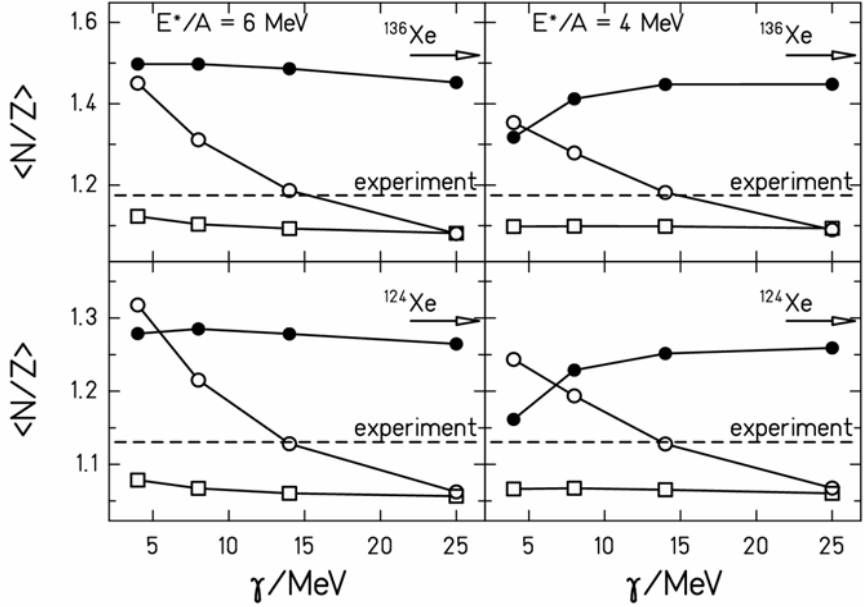

Fig. 4. The <N/Z> ratio versus the symmetry-energy coefficients γ for fragments with Z=10-13 obtained from SMM calculations for $^{136}$Xe (top panel) and $^{124}$Xe (bottom panel) sources, and with excitation energies of 6 A MeV (left panel) and 4 A MeV (right panel). The solid and open symbols represent hot and cold fragments, respectively; for cold fragments: the new evaporation (open circles), the old evaporation (open squares). The dashed line shows the experimental <N/Z>, the arrows indicate the N/Z of the projectiles.

**Neutron fraction of fragments**

Fig.4 shows the SMM calculated <N/Z> ratio of fragments with Z = 10-13 versus γ with the notation as in Fig.3. As was previously demonstrated [42,43], the neutron content of hot primary fragments may increase in multifragmentation region. This is caused by superposition of several factors: The fragment symmetry energy favours formation of symmetric fragments, on the other hand, the Coulomb forces and the entropy of neutrons accumulated in clusters lead to neutron-rich nuclei. The neutron-richness of primary fragments increases usually with decreasing γ. However, if the system contains a big fragment, as in the case of the 4 MeV per nucleon excitation energy (on the right panels of Fig.4), this big cluster contains more neutrons, and the fraction of neutrons in hot fragments with Z=10-13 can drop. It is seen that the primary neutron-rich fragments lose several neutrons during the secondary deexcitation. However, if the evaporation includes the symmetry-energy evolution, the losses are essentially smaller, and the final fragments remain neutron rich. This effect has a simple explanation: Using the experimental masses at all steps of the evaporation, we suppress the emission of charged particles by both the binding energy and the Coulomb barrier, whereas, in the case of small γ at the beginning of the evaporation the binding energy essentially favors the emission of charged particles. When the nucleus cools down sufficiently to restore the normal symmetry energy, the remaining excitation is rather low (i.e. below 1 A MeV) to allow for evaporation of many neutrons. One can see that the new evaporation version predicts γ ≈ 14-15 MeV, in order to describe the experimental <N/Z> for both projectiles.

On the other hand, by considering the α-parameters of the isoscaling (Fig.3), we have seen that the same new evaporation calculation leads to α-parameters which are consistent with the experimental value only at γ ≈ 11-12 MeV. These results are consistent with the physics expected in this case, and the obtained trends are quite reliable, though, they have a qualitative significance only. Taking into account the uncertainty of the model, we can conclude about an essential decrease of the symmetry energy of hot light fragments down to $\gamma \approx$ 11-15 MeV. In this respect, the experimental $\gamma_{ap}$ may be even larger than the real $\gamma$ of the fragments in the freeze-out.

In this work we can not compare the charge partitions calculated by the SMM with experimental data since FRS measures only inclusive yields. However, previously SMM has described very well all charge characteristics, including thermodynamical ones, in similar reactions of fragmentation of relativistic spectators [16,22,49]. As was mentioned, in our calculations we did not find any essential influence of sizes of the thermal spectators on the analyzed isospin characteristics, which are important for the charge partition analysis [16]. Therefore, our model analysis of isospin characteristics is quite reliable. An explanation of the reduction of the symmetry energy in hot fragments is a challenge for future microscopic theories. However, it looks that the most reasonable one would be a considerable decreasing of baryon density of fragments in hot environment [50], which is also consistent with dynamical simulations. Some authors, see ref. [51], have discussed that the observed isospin effect may be related to the influence of the surface symmetry term. Unfortunately, they did not provide a consistent analysis of multifragmentation data in support of their assumption, and, presently, we are not aware of any such an analysis. On the other hand, as was shown in ref. [35], using the Myers-Swiatecki mass formula with the surface symmetry term does not result in a better description of the experimental data. As we discussed, the sub-division of the symmetry energy into bulk and surface terms may not exist at multifragmentation. Moreover, as clear from Fig. 4, the standard evaporation code [45], which uses in nuclear mass formulas both bulk and surface symmetry terms can not explain the observed *<N/Z>* ratio at any symmetry energy of hot nuclei. An importance of the secondary de-excitation process is quite obvious, since the primary fragments can have an excitation energy up to 2-3 MeV/nucleon [52], and the evaporation changes essentially their isospin composition. We emphasize that besides the excitation energy this process can depend on initial properties of hot fragments. In the case of nuclear multifragmentation such a secondary de-excitation can lead to important consequences in the final isotope yield, which can be used for extraction of the symmetry energy of these fragments.

**Conclusion**

In this work, the recent FRS data on fragment isospin were analysed within the statistical multifragmentation model. We have found a decrease of the symmetry coefficient of primary fragments, which are formed at the freeze-out stage. This is consistent with the other investigations of the symmetry energy of hot light nuclei in multifragmentation [10,39,53,54,55,56,57]. In the present study we use both, the *<N/Z>* ratio and the isoscaling for the purpose of investigating the symmetry energy coefficient. The coincidence of both methods makes us more confident about the decrease of the symmetry energy of hot light fragments. Our result shows that the properties of light nuclei can change during the nuclear liquid-gas phase transition. With the multifragmentation reaction one can investigate hot nuclei produced at temperatures around $T \sim$ 3-8 MeV, and at densities of the matter around $\rho$ = (0.1 - 0.3) $\rho_0$. As was demonstrated in [3], the decrease of the symmetry energy in hot fragments may have important consequences for supernova processes, where densities and temperatures close to the nuclear multifragmentation case can be reached.

One of the authors (A.S.B.) thanks GSI for warm hospitality and support. We thank F. Gulminelli, W. Trautmann, and M. B. Tsang for illuminating discussions. The experimental analysis in this work forms part of the PhD thesis of D. Henzlova.

---

[1]    P. Donati et al., Phys. Rev. Lett. **72**, 2835 (1994).